# PRODUCTION OF ACTIVATED AND GRAPHENE-LIKE CARBON MATERIALS FROM RICE HUSK

Z.A. Mansurov[1,2], G.T. Smagulova[1,2], F.R. Sultanov[1,2], B.B. Kaidar[1,2], Z. Insepov[3,4]
A.A Imash[1,2*]

[1]Instute of Combustion Problems (ICP), Almaty, Kazakhstan
[2]Al-Farabi Kazakh National University (KazNU), Almaty, Kazakhstan
[3]Purdue University, West Lafayette, 500 Central Drive, IN USA
[4]Satpayev Kazakh National Technical University, Satpayev 22, Almaty, Kazakhstan
E-mail: ZMansurov@kaznu.kz
[*]Corresponding author: A.A. Imash, iimash.aigerim@gmail.com

**ABSTRACT:** The main research efforts are aimed at finding inexpensive materials that can be converted into nanostructured carbon-containing materials, such as activated carbon and graphene. Rice husks are one such material, especially in developing countries, where more than 95% of Rice husks are produced worldwide. Although numerous studies have been conducted on the production of activated carbon and graphene structures from Rice husk, the existing scientific information is still widely scattered in the literature. Therefore, this review article provides extensive information on made by various researchers, including the Institute of combustion problems (Almaty, Kazakhstan), regarding the production of nanostructured carbon-containing materials from rice husk and its adsorption characteristics. The properties and pre-treatment of rice husk in relation to the production of activated carbon and multilayer graphene are discussed. The activation of rice husk by physical and chemical methods under different conditions is considered.

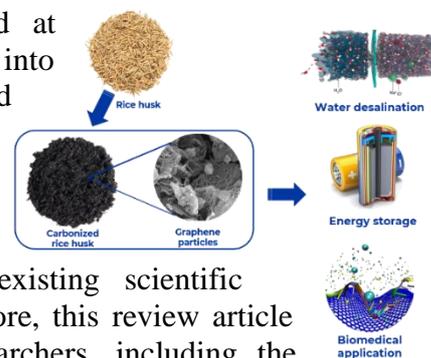

## INTRODUCTION

Porous carbon materials are the most in-demand product among carbon materials. The creation of nanostructured carbon-containing materials, including activated carbons, is one of the priority research areas rapidly developing in the Commonwealth of Independent States (CIS) countries and all over the world. Due to their unique textural characteristics and high specific surface area, they find application in a wide variety of practical applications.

This review collected the result of scientific work in the field of processing of Rice husk (RH) to produce porous carbon materials, activated carbon and their application as an electrode material of energy storage systems, water and wastewater treatment, water desalination, etc.

With the discovery of graphene structures scientists associate the solution of many problems and the search for new materials with improved performance characteristics. The problem of large-scale production of graphene and graphene-containing materials restrains the mass application of these types of materials. The second section of the review presents the current achievements in the synthesis of graphene and graphene-containing materials from Rice husk.

Institute of combustion problems for many years has been working on the issue of synthesis of activated carbons from various materials and their application. These works were initiated by Doctor of Chemical Sciences, Professor R.M. Mansurova and continue to this day by her students.

In 2001, R.M.Mansurova defended her doctoral thesis on the topic: "Thermocatalytic and thermal synthesis of carbon-containing sorbents, refractories, nonflammable compositions and catalysts" on the specialty 01.04.17 – "Chemical physics, including the physics of combustion and explosion".

Prospects of research on the study of nanostructured carbon sorbents and the development of technology for their production in Kazakhstan from local raw materials, launched by Professor R.M. Mansurova, raises no doubt, because they are associated with the technology of mining and processing of various materials, environmental, medical and many other problems. Their continuation will be the best expression of appreciation and respect for the memory of their Teacher - Doctor of Chemical Sciences, Professor Raushan Magzumovna Mansurova.

# 1. ACTIVATED CARBON FROM RICE HUSK

Activated carbon (AC), known to humanity since ancient times, appears to us as a completely unique material due to its properties. AC are usually characterized by a large surface area, high microporosity and adsorption capacity, which allows them to be used as effective adsorbents in water purification processes, for solvent extraction, gas separation, catalysis, etc. [1, 2]. Scientists are still discovering new facets of the properties and applications of activated carbon. Activated carbon theoretically can be produced from any carbon-containing material rich in elemental carbon.

Technologies for obtaining activated carbon from natural fossils, such as charcoal [3] and hard coal [4], from wood species [5], as well as methods of obtaining from agricultural waste and biowaste are well known: straw, cake, bamboo, cotton residues, nut shells, fruit pips, fruit seeds, fruit peels, coconut shells [6], olive pips, sunflower oil residues, corn cobs [7], tropical peat soil [8] and from more exotic materials such as coffee [9] and tea residues [10], rotten strawberries [11], and even wine-making wastes [12]), also polymers [13] can serve as precursors. Rice husk [14], which is a multi-tonnage renewable source for the production of AC, occupies a special place among the starting materials for AC production. Textural characteristics (specific surface area, pore diameter and volume, pore size distribution, wall thickness, size of structural elements, etc.), mechanical (abrasion resistance), electrical (electrical conductivity), and chemical properties (functional groups, types of chemical bonds, carbon, oxygen, and other elements content) of activated coals depend on:

1. source materials (listed above);

2. pre-treatment conditions (if any);

3. carbonation conditions: temperature, carbonation time, atmosphere, gas flow rate, as shown in [15];

4. physical activation conditions (if available) [16];

5. conditions of chemical activation (if any) as shown in [17].

In recent years, there has been a growing interest in the production of activated carbon from agricultural by-products and waste [18]. The agricultural by-products available in large quantities are cake, a fibrous by-product produced by grinding sugarcane and rice hulls [19]. Annual global rice production is estimated to be about 571 million tons, yielding about 140 million tons of Rice husk available for use per year [20]. Despite the widespread use of Rice husk as fuel for plant boilers, power and steam generation, animal feed, or as raw material for paper and cardboard production, residues are still in surplus, which creates a disposal problem. Obtaining porous carbon materials with good adsorption properties from Rice husk allows, on the one hand, to solve the problems of disposal of these wastes, on the other hand, it makes it possible to obtain good materials - sorbents for drinking water and wastewater treatment.

Rice husk consists mainly of hemicellulose, cellulose, and lignin, as well as $SiO_2$ [21-44]. The bulk density of rice hulls is low and is in the range of 90-150 kg·m$^{-3}$. Obtaining activated carbon and porous materials largely depends on the method of synthesis and conditions of rice husk processing [25].

In [26], researchers obtained 2D porous carbon structures from Rice husk by environmentally friendly and recyclable salt activation. The authors used a mixture of NaCl/KCl salts, which are templating agents for the formation of mesopores, but also necessary for the formation of 2D carbon structure. According to the results of the work, porous carbon sheets have a large surface area and many open micro- and mesoporous located on a flat carbon sheet. The porous carbon material showed a high specific capacitance (288 F·g$^{-1}$ at 0.5 A·g$^{-1}$.), due to the effective use of pores, the performance and cyclic stability of the electrode material. In [27, 28] a group of researchers obtained highly porous materials from activated carbon with micro/mesoporosity by carbonization of rice husk and further treatment with $K_2CO_3$. The authors investigated the microstructure and characteristics of rice husk-derived activated carbon (RHAC). The samples obtained at 600°C and 800°C are designated as RHAC-600 and RHAC-800.

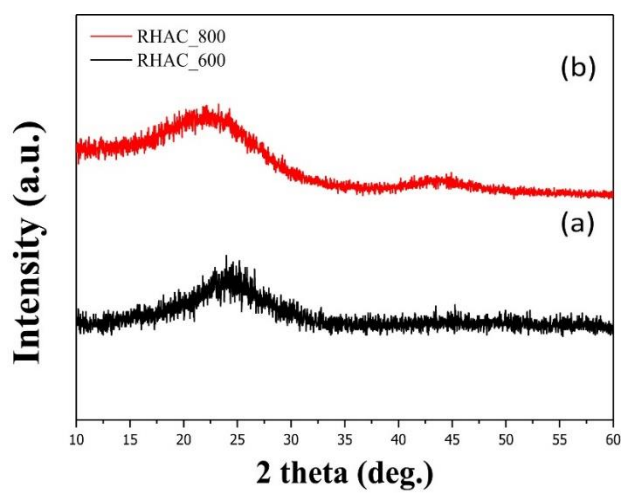

Figure 1. X-ray diffraction patterns of (a) RHAC-600 and (b) RHAC-800 samples [27].

Figure 1 shows the X-ray patterns of samples RHAC-600 and RHAC-800. Two typical diffraction peaks at 2θ 22.5° and 43° can be attributed to reflections from crystalline planes (002) and (110) of graphite, while the broad peaks indicate an amorphous structure.

The sample RHAC-800 has an amorphous structure with a large surface area ($S_{BET}$) of 1583 m$^2$·g$^{-1}$ and a pore volume of 0.93 cm$^3$·g$^{-1}$. The work is generally aimed at obtaining composite materials for lithium-sulfur (Li-S) batteries. The researchers placed elemental sulfur in the micropores using the solution infiltration method to form RHAC@S-based composites. Freshly prepared RHAC@S composites with sulfur contents of 0.25 mg·cm$^{-1}$ and 0.38 mg·cm$^{-1}$ were tested as cathodes for Li-S batteries. These results demonstrate that porous materials are of interest as cathode materials for the development of high efficiency Li-S batteries. The authors believe that these results will open up new opportunities for the development of high-performance Li-S batteries using efficient porous carbon materials.

Another method for obtaining porous carbon from Rice husk: the self-template method for obtaining anode material.

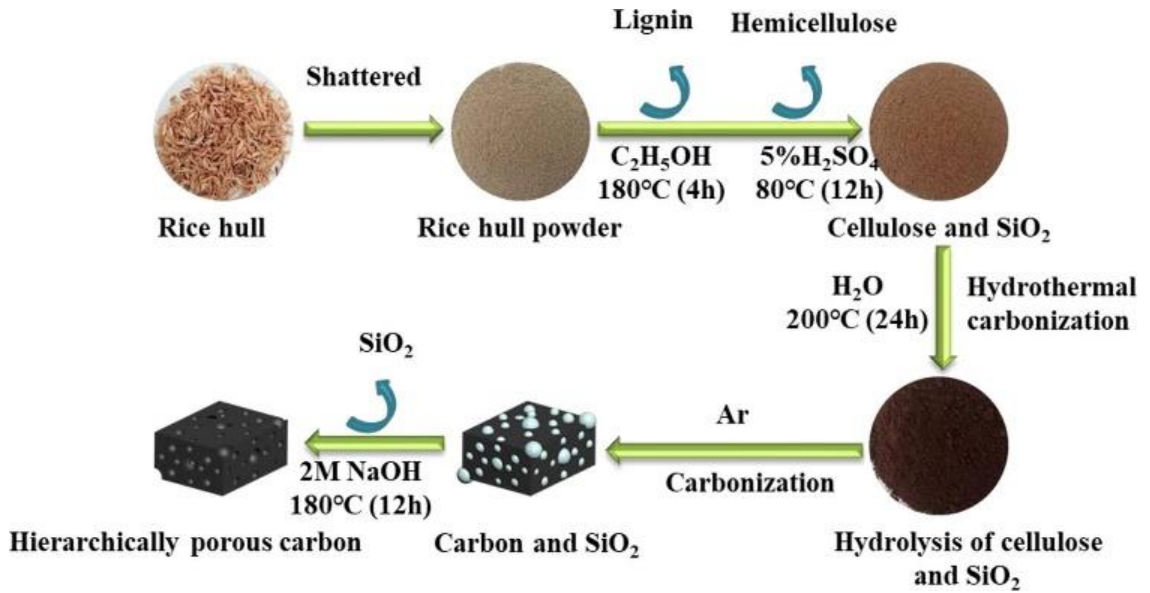

Figure 2. Cycle of the process of obtaining porous carbon from Rice husk [29].

Jiazi Hou et al. used $SiO_2$ as a template to produce porous carbon materials from environmentally friendly Rice husk. The porous carbon material subjected to hydrothermal treatment for 24 hours exhibits excellent micromorphology and excellent electrochemical properties. The specific charge capacity can reach 679.9 mAh·g$^{-1}$ after 100 cycles at a current density of 0.2 A·m$^{-2}$. This significant improvement in electrochemical performance is attributed to the edge structure, defects, and the large specific surface area that the porous structure possesses [29].

The success of the adsorption process is determined by the behaviour of the adsorbent before and after the adsorption, equilibrium is reached. Thus, the ideal carrier in the general case is one in which the easy accessibility of the surface is combined with its development. The number, size, length and shape of the pores mathematically determine both of these parameters. In the case of micropores, due to the relative proximity of the pore walls, the interaction potential and orienting effects are much greater than in the case of wider pores. Capillary condensation phenomena are often observed in mesopores, while macropores significantly reduce the role of diffusion processes and are described by relatively simple equations. The hierarchical structure of adsorbent micropores is shown in Figure 3a.

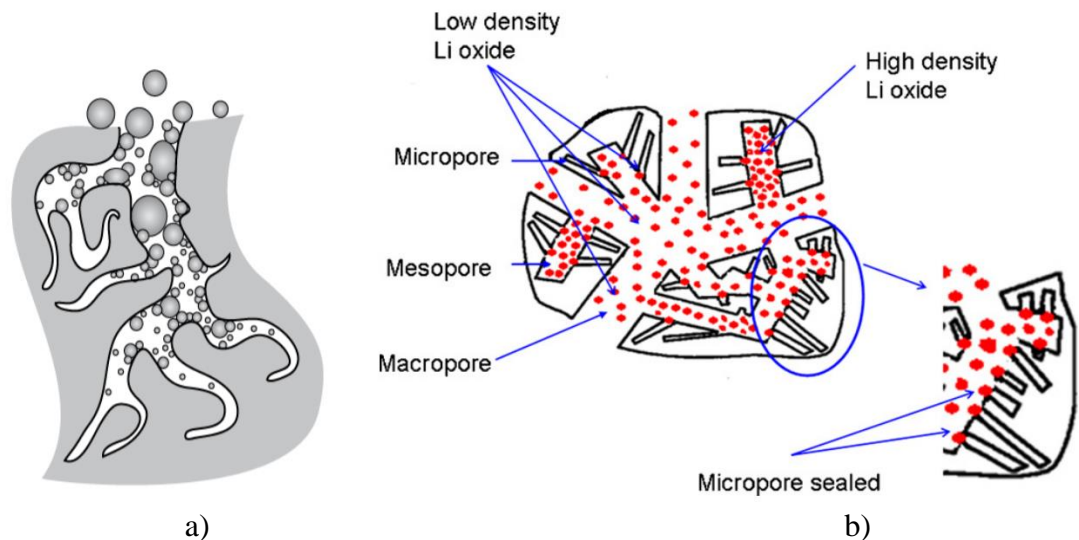

a)            b)

Figure 3. a) The system of macro- and micropores providing a developed

adsorption surface, as well as a certain size selectivity of the adsorbed molecules [30, 31], b) Placement of Li oxides in pores of different sizes [32].

It should be noted that the micropores are sealed immediately. This seems to be the reason for the inequality of adsorbed and desorbed substances as shown (in Fig. 3b) in the research work [32].

Alessandra P. Silva et al. obtained activated carbon from rice husk at different carbonization temperatures, and found that at lower carbonization temperatures (450 to 650°C), the activated carbon electrode from RH showed a higher desalination performance with a salt removal efficiency of 15.5 mg·g$^{-1}$ (1.2V and $C_0$ = 600 mg·l$^{-1}$). The researchers suggest that the optimal carbonization temperature is 600°C, explaining that the electrodes remain stable up to 80 cycles of electrosorption/desorption, with kinetics and electrosorption capacity remaining close to the original values. This activated carbon rice husk electrode can be applied in capacitive deionization. This industry has attracted much attention as a promising desalination technology because inexpensive carbon electrodes can be used on a large scale [33].

Songlei Lv et al. in their research obtained activated carbon from Rice husk by pyrolysis of Rice husk, with simultaneous KOH activation and modification with tetrasodium salt of ethylenediaminetetraacetic acid (EDTA-4Na) to adsorb phenol from water. In this work, rice husk was carbonized at 500°C and activated at 750°C with a mass ratio of carbon to KOH of 1:3, and an activated carbon with large microporosity and a specific area of 2087 m$^2$·g$^{-1}$ was obtained; the maximum phenol adsorption for AC is 194.24 mg·g$^{-1}$ [34]. In [35], the authors fabricated nanoporous coals from rice husk through KOH activation and investigated the reactions occurring during the production of porous carbons based on alkaline activation from pure carbon materials. Figure 4 shows the two-step mechanism of pore formation.

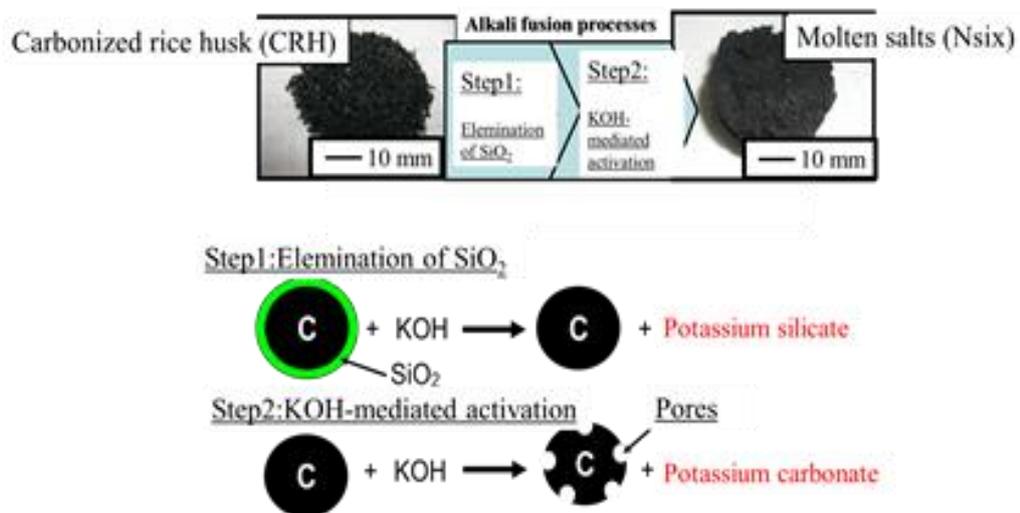

Figure 4. Prospective treatment by alkaline melting (KOH activation) of carbonaceous Rice husk [35].

Based on reactions describing in [35] the effects of KOH on the removal of $SiO_2$ from the surface of Rice husk, the pore-forming process was evaluated for the first time. The micropore volume of NSi70 was determined as 0.385 cm$^3$·g$^{-1}$ and the corresponding nanoporous carbon (NPC) was obtained as 1.778 cm$^3$·g$^{-1}$. Micropore volume per volume was evaluated by reactions in works. The true NPC density was assumed to be 2.0 g·cm$^{-3}$.

・ Amorphous carbon: 0.512·12·0.385·(0.5)$^{-1}$ = 55.44 cm$^3$
・ NPC: 0.5·12·4·1.778·(0.5)$^{-1}$ = 85.34 cm$^3$

The volume of 12 mol of amorphous carbon was taken to be 72 cm$^3$, and 4 mol of NPC was defined as 24 cm$^3$. Thus, the authors state that the difference in micropore volume per unit volume was 28.98 cm$^3$. Thus, the amount of carbon removed during pore formation can be estimated as follows.

・Microporation: 100% ·28.98/48 = 60%

・Macro- and mesopore formation: 100% − 60% = 40%

According to the above estimates, 60% of carbon was removed during microporation, and 40% of carbon was removed during the formation of macro-/mesopores [35].

The structure and composition of Rice husk play a very important role in the synthesis of activated carbon. Rice husk contain 15 to 20 wt. % silica and other inclusions. Silicon dioxide in rice husk affects the formation of mesopores during activation of KOH, which additionally affects the electrochemical characteristics of the resulting activated carbon. In [36], the authors examined the effect of $SiO_2$ on the porosity of activated carbon. When $SiO_2$ is not removed from the rice husk, the formation and increase of mesopores stops during activation. The resulting activated carbon has a high microporosity ratio, high specific area (up to 3263 m$^2$·g$^{-1}$), high specific capacitance (315 F·g$^{-1}$ at 0.5 A·g$^{-1}$), but low capacity (51.7% capacity retention when current density is increased from 0.5 A·g$^{-1}$ to 20 A·g$^{-1}$). When $SiO_2$ is removed from the rice husk, the pores formed by $SiO_2$ removal are etched to mesopores during activation; the resulting activated carbon exhibits a high mesoporosity ratio, a relatively high specific surface area of 2804 m$^2$·g$^{-1}$ and a high capacity of 278 F·g$^{-1}$ at 0.5 A·g$^{-1}$ with excellent capacity (capacity conservation of 76.6% with increasing current density from 0.5 A·g$^{-1}$ to 20 A·g$^{-1}$) (Figure 5).

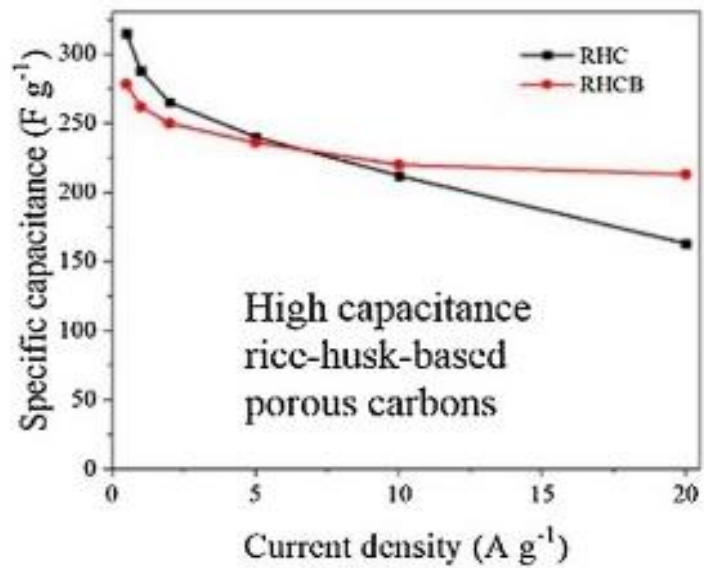

Figure 5. The dependence of the specific capacitance on the current density of porous carbon based on rice husk (carbon materials by removing the silica RHCB and carbon material without removing the silica RHC) [36].

The uniqueness of activated carbon is due to the fact that the AC can be considered as a carbon-carbon composite material in a certain approximation due to the fact that it can contain: particles with ordered graphitized structure, as well as with disordered porous structure, multilayer and single-layer graphene, carbon fibers and tapes, hollow spherical structures and even multilayer carbon nanotubes. The combination of these particles gives the variety of properties of activated carbon materials that we know.

## 2. RESEARCH OF THE INSTITUTE OF COMBUSTION PROBLEMS TO PRODUCE CARBON SORBENTS BASED ON RICE HUSK

*2.1 Rice husk-based sorbents for water treatment from oil pollution*

A nanostructured carbon-mineral sorbent based on carbonized plant raw materials containing carbon and silicon oxide was synthesized in the R.M. Mansurova Laboratory of Carbon Nanomaterials of the Institute of combustion problems.

In [37, 38] carbon-containing materials obtained as a result of carbonization of plant raw materials such as apricot and grape seeds, walnut shells, cane and poplar wood were studied.

The obtained sorbents were studied for their absorption capacity with respect to petroleum products. To determine the absorption characteristics of the sorbent, 500 ml of water, 100 ml of machine oil were poured into a flask and carbonized Rice husk (CRH) was added. After 20 minutes, the carbonized RH completely absorbed petroleum products and the water surface became clean.

In this work, the adsorption characteristics of the CRH with respect to oil were determined. It was found that rice husk (carbonized at 600-700°C) with a mass of 0.1 g sorbs >12.0 g of machine oil and <1.5 g of water, which indicates the applicability of CRH as a sorbent for cleaning oil spills.

The effect of the carbonization temperature of CRH on the amount of adsorbed oil was also investigated. Figure 6 shows the dependence of the amount of light and machine oil adsorbed by 1g of CRH depending on the temperature of carbonization of RH. The maximum amount of adsorbed oil is observed for the sample of CRH obtained at a carbonization temperature of 600-700°C - 7 g for light oil and 12 g for machine oil.

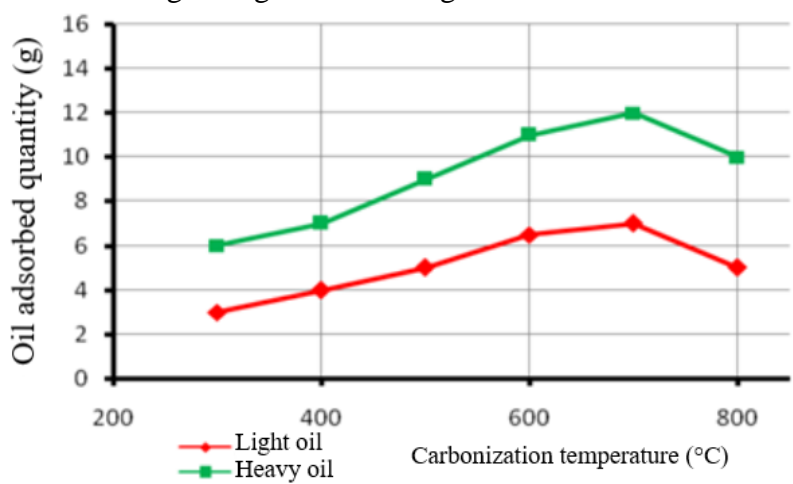

Figure 6. Dependence of sorption properties of CRH on the carbonization temperature for light and machine oil [39].

In the work the influence of the density of petroleum products on the sorption capacity of sorbents made of carbonized rice husk and apricot stone was studied (Fig. 6). As the density of petroleum products increases, the sorption capacity increases linearly. For sorbents from apricot stone sorption characteristics for different densities of petroleum products changes insignificantly.

For the studied range of densities, the sorption capacity of CRH increases in 3 times, while for ASC (apricot stone carbonization) it increases in 8 times. The sorbent obtained by carbonization of rice husk shows a higher adsorption efficiency of various petroleum products compared to other sorbents.

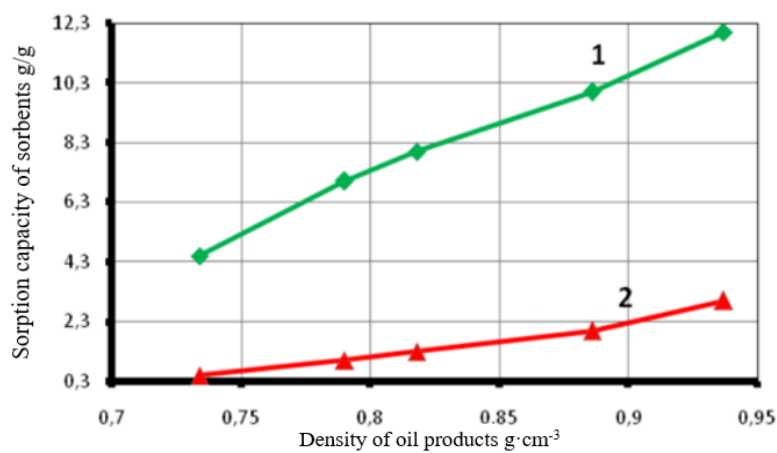

Figure 7. Dependence of sorption properties of carbonized rice husk (1) and carbonized apricot stone (2) on oil products density [39].

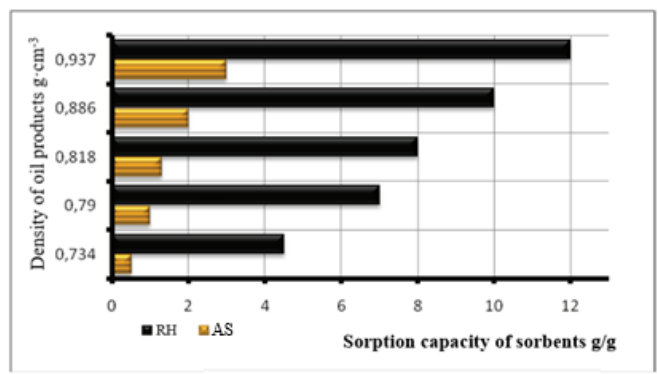

Figure 8. Sorption characteristics of carbonized rice husk and carbonized apricot stone as a function of petroleum product density [39].

The authors showed [38] that the sorbent obtained by carbonization of rice husk shows high productivity (Figure 8) and opens up the possibility of its practical application for removal of spilled oil and petroleum products from the surface of water areas [39].

### 2.2 Extraction of fusicoccin on a carbon-mineral sorbent

In [40, 41] studies on the extraction of fusicoccin on a sorbent from rice husk were presented. The Italian scientist Alessandro Ballio as a phytotoxin of the phytopathogenic fungus Fusicoccum amygdali Del. discovered Fusicoccin in 1964. [42]. Fusicoccin exhibits a variety of physiological and biochemical properties. It is known that the problem of isolating preparative amounts of natural fusicoccin from higher plants is an extremely difficult experimental task. In this connection, the problem of separating biologically active substances arose. To solve this problem, we used the liquid chromatography technique using a sorbent made of rice husk. For a comparative analysis of the specific characteristics of the sorbent made of RH was used organic gel octylsepharose 4B-CL (Sweden), used in the world practice now. The results of the study are presented in Figure 9.

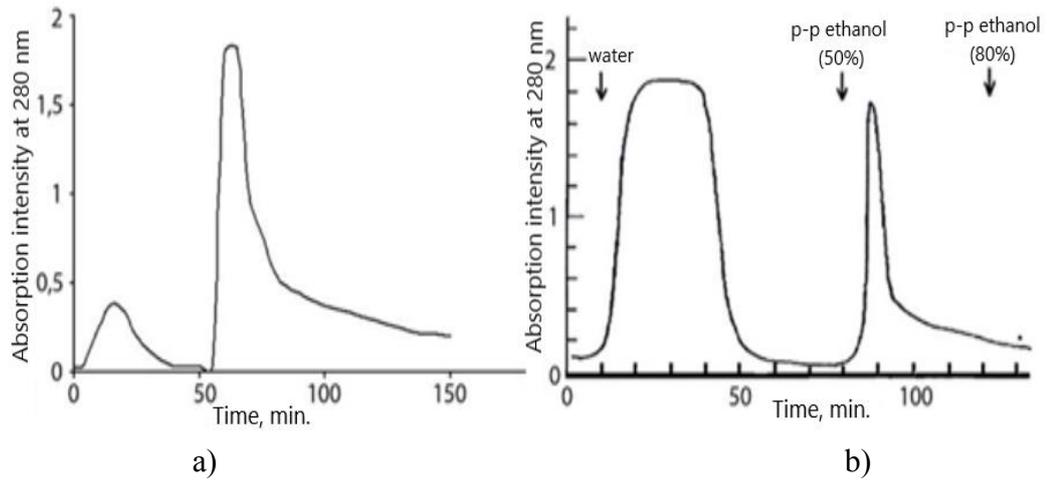

a) b)

Figure 9. Fusicoccin chromatographic separation curves on column with CRH-750 (a) and octylsepharose 4B-CL organic gel (b) [43].

The authors of the study showed that when using a nanostructured sorbent based on RH, the separation process takes less time [43].

### *2.3 Study of carbonized Rice husk by EPR method*

In the article [44], the authors present the results of the study of carbonized samples of Rice husk by the EPR method (Figure 10).

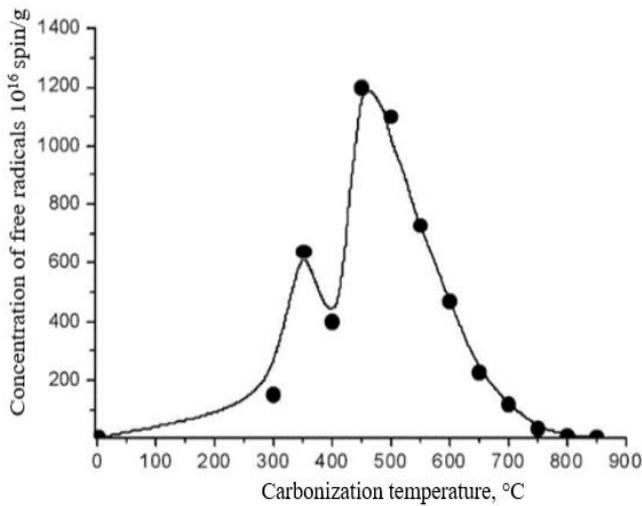

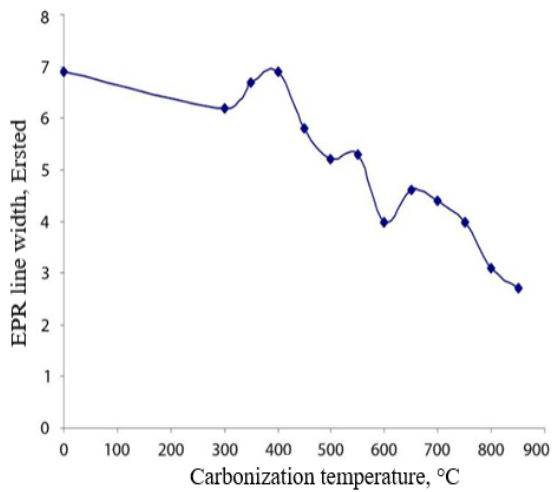

a) b)

Figure 10. a) Influence of carbonization temperature on the concentration of free radicals in RH, b) Dependences of the EPR line width of RH samples on the carbonization temperature [44].

Figure 10 (b) shows that the width of the ΔH line gradually decreases to 2.7 Ersted (850 °C). However, in the temperature range 550-650 °C there is a "maximum-minimum" transition with the following values of ΔH: 550°C - 5.3 Ersted, 600°C - 4.0 Ersted, 650°C - 4.6 Ersted. The authors suggested that at temperatures corresponding to the minimum concentration of free-radical states, especially at its high values, formation of various carbon nanostructures takes place. This is explained by the fact that such nanostructures have diamagnetic properties. This should, along with the formation of graphene-graphite-like structures, lead to a sharp drop in the EPR signal intensity. The conducted EPR studies of scientists confirm the formation of nanostructures in carbonized Rice husk.

### 2.4 Sorption of gold (III) on carbonized Rice husk

In [45] studied the sorption of gold (III) on sorbents CRH-1, CRH-2 and CRH-3 in 0.25 n HCl solution using classical and electrochemical techniques. The results showed (Fig. 11) that the best 100% sorption of gold (III) has a sorbent CRH-1.

The results showed (Fig. 11a) that the best 100% sorption of gold (III) has a sorbent CRH-1. The sorbent CRH-2 for the sorbent CRH-2 for a period of 1 hour sorbs about 60% of gold (III), and the sorbent CRH-3 for the same time sorbs only 20% of gold (III). Sorbent CRH-3 sorbs only 20% of gold (III). In the case of sorbent CRH-1 complete almost 100% sorption occurs within 5 minutes. The nature of the course of kinetic curves shows that the rate of sorption of gold (III) decreases in the series: CRH-1 > CRH-2 > CRH-3.

Table 1. Potentials of carbonized sorbents in 0.25n HCl solution.

| Sorbent | E, V (Silver chloride electrode SCE) |
|---|---|
| CRH-1 | 0.229 |
| CRH-2 | 0.362 |
| CRH-3 | 0.376 |

The restorative capacity of the sorbents of the CRH series is confirmed by the measured values of the steady-state potentials of these sorbents. The results of measuring the potentials are presented in Table 1.

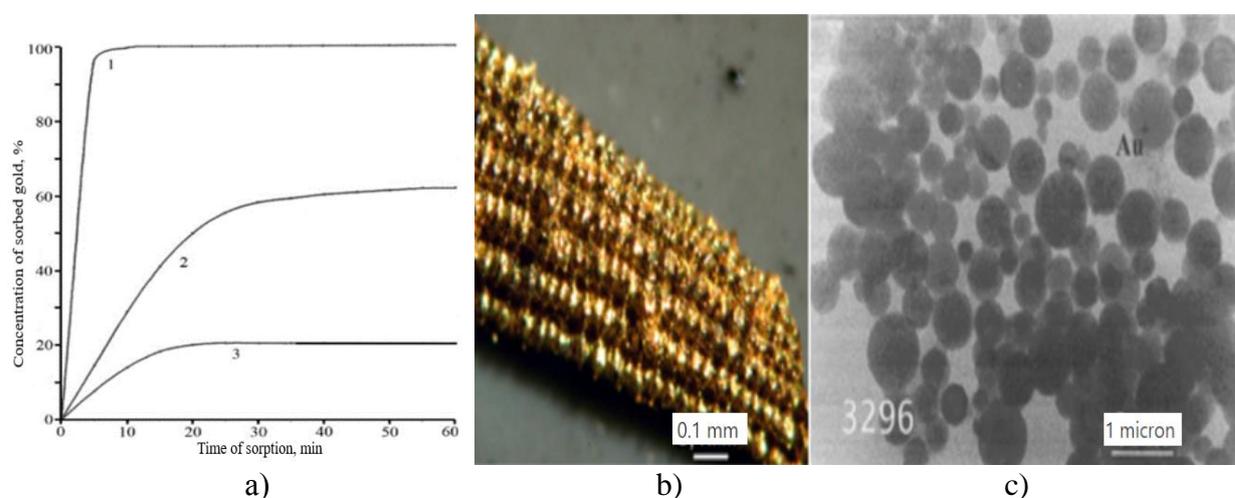

a)      b)      c)

Figure 11. a) Kinetic curves of gold (III) sorption on different sorbents (1-(CRH-1), 2-(CRH-2) and 3-(CRH-3)); b) Optical microscopy micrograph; c) Electron microscopy photo of CRH-1 after sorption [45].

The sorbents of CRH are not only ion-exchange but also redox sorbents. From the data on potentials it follows that the potential difference between the oxidizing gold (III) and reducing sorbents CRH-1, CRH-2 and CRH-3 is 0.51, 0.38 and 0.36 V, respectively. As described above, for practically complete (99%) course of any redox reaction a difference of real (steady-state) potentials equal to 0.24 V is required. In the case of CRH-1 the potential difference is the largest, so this sorbent should be easier to recover gold (III).

It is established that the most optimal sorbent is CRH-1, and the dynamic capacity of CRH-1 for gold is 10.5 mg·g$^{-1}$.

Studies on the extraction of noble metals on activated carbons from Rice husk are continuing now. As part of the project "AP05134691" "Development of a method for electrochemical concentration of noble metals using nanoporous electrode materials." (2018-2020), comprehensive studies were carried out on the electrochemical extraction of noble metals gold and silver from their solutions, at different values of solution pH, concentration and the presence of impurity ions [46, 47]. Physical and chemical characteristics of activated carbons from Rice husk were investigated. According to the results of BET analysis, the value of specific surface area for activated carbon from RH is 2817.70 m$^2$·g$^{-1}$, the specific pore volume is 1.59 cm$^3$·g$^{-1}$ and the average pore diameter is 2.61 nm. The electrode for electrochemical deposition of gold ions was made based on the obtained carbon material. Studies were conducted to determine the optimal pH value of the aqueous solution containing gold ions; it was found that the optimal pH value for the electrochemical sorption of gold is 1.39. In this work, the laws of precipitation and dissolution of gold ions (III) from its solutions have been studied by cyclic voltammetry. Diffusion coefficient for gold in aqueous medium was calculated as 2·10$^{-6}$ cm$^2$·s$^{-1}$. It was determined that as a result of electrochemical sorption nanoparticles of gold with particle size ~100 nm were deposited on the electrodes made of carbon material from RH (Fig. 12).

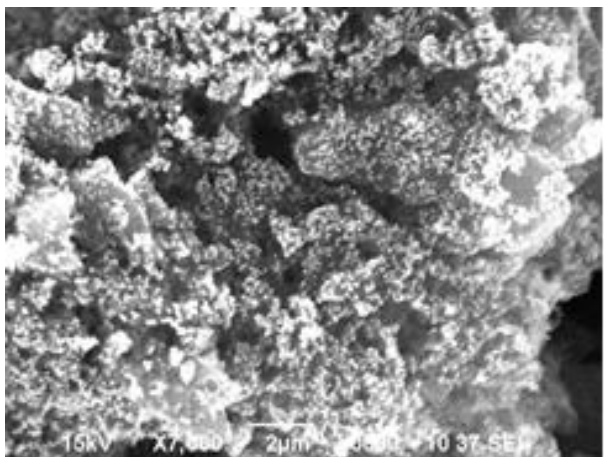

Figure 12. SEM image of the surface of the carbon electrode from RH after gold sorption [46].

The efficiency of different eluate solutions on desorption of gold (III) ions from the sorbent surface was also studied (Table 2). The use of a mixture of acetone with water and 1 g/l NaOH for the extraction time of 2 hours provided a degree of desorption of gold 88%, and the use of a mixture of isopropyl alcohol with water and 1 g/l NaOH for the process duration of 5 hours - 67%.

Table 2 - Values for the desorption efficiency of gold into the desorbing solution under different conditions [46].

| Sorbent | Eluate (desorbent liquid) | NaOH, g/l | T, K | Desorption rate Au, % | Retrieval time Au, h |
|---|---|---|---|---|---|
| Rice husk carbon material with | Acetone+water+NaOH | - | 303 | 35 | 2 |
| | | 1 | | 88 | |
| | | 2 | | 96 | |

| gold content (0.01 g Au/1 g carbon) | | 10 | | 93 | |
| --- | --- | --- | --- | --- | --- |
| | | 20 | | 91 | |
| | Isopropyl alcohol + water+NaOH | - | 303 | 15 | 5 |
| | | 1 | | 67 | |
| | | 2 | | 74 | |
| | | 10 | | 76 | |
| | | 20 | | 77 | |

As can be seen from Table 2, the most complete dissolution of gold from the sorbent surface, and consequently, the regeneration of activated carbon is observed when choosing mixtures of acetone + water + 2 g/l NaOH, isopropyl alcohol + water + 20 g/l NaOH as eluate.

Schematic diagram of the electrosorption and desorption system is shown in Fig. 13. The process includes gold leaching with aqua regia solution 3 followed by electrosorption of dissolved gold on carbon electrode 1 (Fig. 13, a) (the anode is a Pt-plate 2).

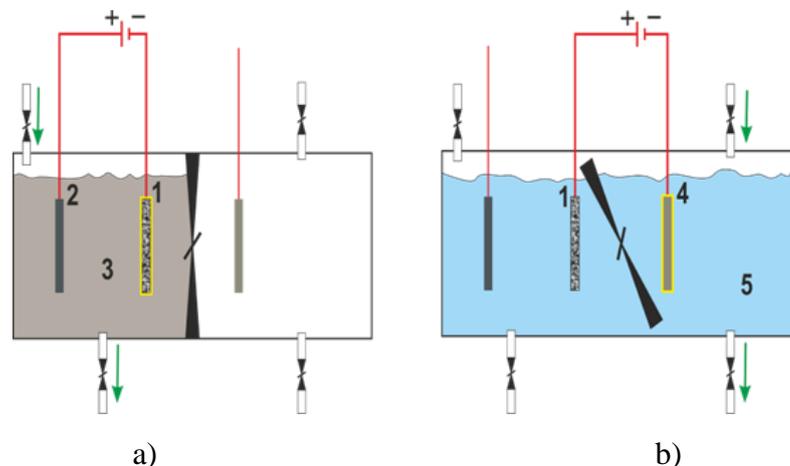

a) b)

Figure 13 - Schematic diagram of an electrosorption and desorption system for gold (III) ions [46].

The authors state that the gold solution after sorption by carbon electrode is returned to leaching circuit and gold loaded carbon is fed into elution zone separated with leaching circuit by movable baffle 5 (Fig. 13, b). Gold was extracted from the gold-loaded electrode 1 using a hot eluate solution (a mixture of acetone and water at a volume ratio of volume ratio of 1 : 1, in which 1 g of sodium hydroxide 5 is dissolved) and electrosorbate on a stainless steel cathode 4 in an electrolytic cell (Fig. 13, b). The paper concludes that the activated carbon-based electrode can be used repeatedly in the process of gold electro sorption from solution and reverse desorption.

In [48] the possibility of obtaining effective bio sorbents for industrial wastewater treatment based on RH was shown. Bio sorbents are systems providing sorption of various metals from dilute solutions (wastewater of industrial enterprises) by immobilized cells of microorganisms with specific sorption activity in relation to metals. The authors showed that the attachment of yeast cells Rhodotorula glutinis var glutinis, and cell culture of bacteria Pseudomonas aeruginosa on a carburized carrier allows to obtain a highly effective bio sorbent, the sorption activity of which with respect to copper and lead ions was 93-95%. The data of the results are shown in Fig. 14.

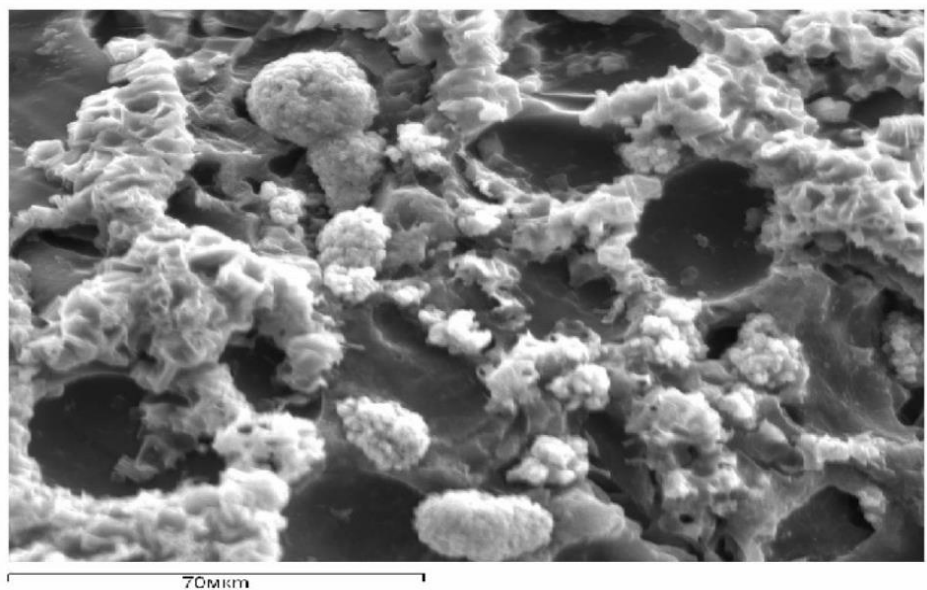

Figure 14. Electron microscopic images of RH carbonized at 650 °C and cells of Rhodotorula glutinis var. glutinis after deposition of Cu ions, (magnification x1000) [48].

In the works carried out by Russian scientists together with scientists from Germany [31], adsorption of haemoglobin on carbonized Rice husk at different temperatures was studied. The presence of free haemoglobin in the blood plasma of patients is extremely undesirable and leads to massive damage of internal organs. Therefore, haemoglobin adsorption is the preferred process. Two varieties of sorbents, T800 (CRH at 800 °C) and T600 (CRH at 600 °C), showed a similar temperature dependence, characterized by an increase in adsorbed substance when higher temperatures are used. The results are shown in Figure 15.

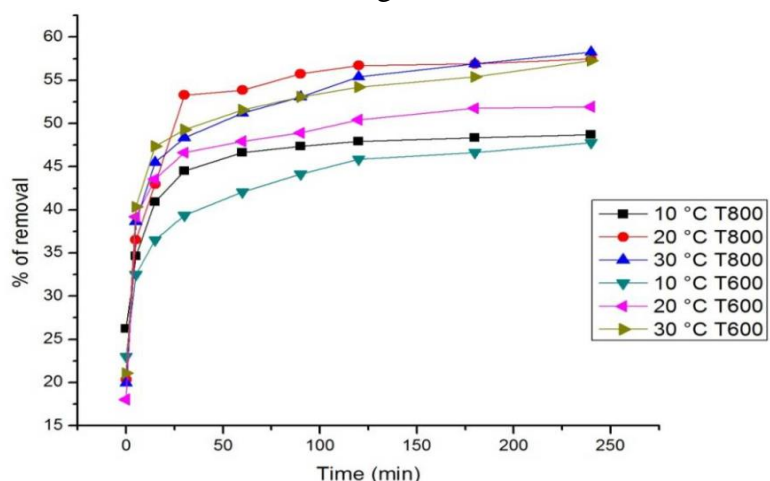

Figure 15. Dependence of haemoglobin adsorption on temperature on sorbents T600 and T800 [31].

More detailed temperature characteristics of the process can be obtained by analysing diagrams (Figure 16) reflecting the dependence of the amount of material adsorbed by the surface on small temperature fluctuations.

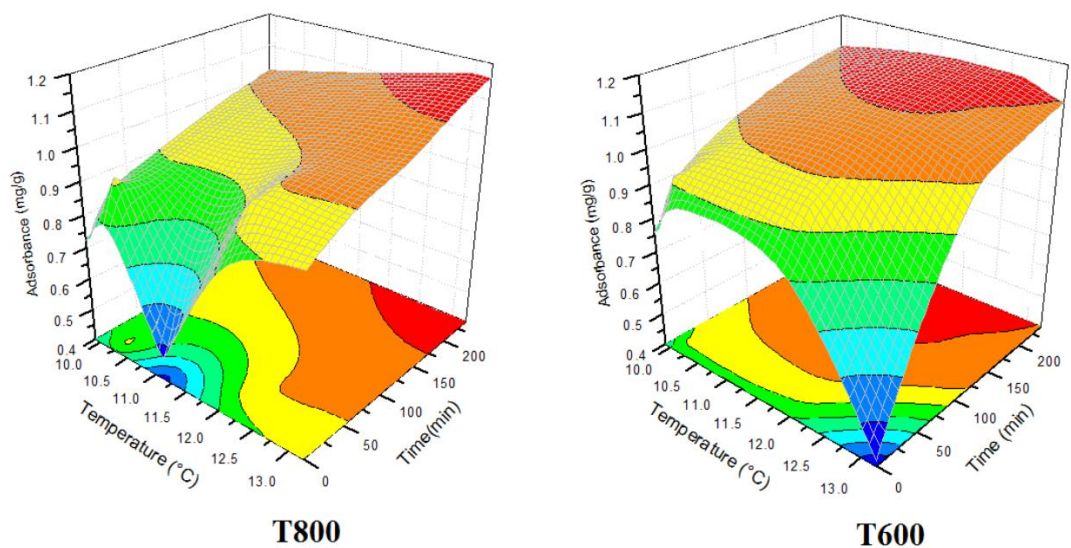

Figure 16. Variations in the kinetic characteristics of haemoglobin adsorption during temperature fluctuations around 30°C [31].

In this case, the positive effect of increased temperature is probably not related to any chemical reactions and their activation energy, but rather to an increase in the diffusion of protein molecules, which leads to a more rapid achievement of the surface located in a relatively closed volume of micropores.

Comparison of the patterns of haemoglobin adsorption on CRH at different temperatures indicates the existence of local minima and maxima of adsorption, as well as the complex nature of the dependence of the degree of adsorption on temperature. In general, experiments with haemoglobin adsorption on XRF led to the conclusion that the extraction of haemoglobin is most active in the area of relatively high temperatures, reaching 60% of the extraction level at a given ratio of the volumes of liquid and solid phases.

By high-temperature carbonization of the initial raw materials at the Institute of Combustion Problems, carburized rice husk (CbRH) was obtained, which also gives the possibility of obtaining probiotic-bio sorbents [49]. The observation of scientists led to the conclusion that the decrease in the population level of Enterobacteriaceae in the intestines of animals receiving probiotics after receiving ampiox may be due to both the stimulation of antimicrobial activity of lactobacilli immobilized on CbRH and more intensive adsorption of enterobacteriaceae lipopolysaccharides (LPS) on the sorbent [50]. In this connection [51], the antimicrobial activity of free and immobilized lactobacilli cells against hospital strains of Enterobacteriaceae was determined. Cultures of 8 bacterial species belonging to the family Enterobactericeae were used as target microorganisms in this case (Fig. 17).

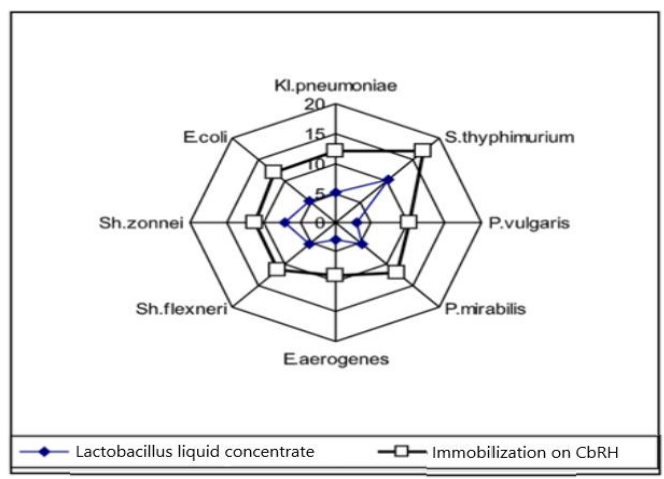

Figure 17. Antimicrobial activity of free and immobilized Lactobacillus cells on CbRH against Enterobacteriaceae [51].

It turned out that the antimicrobial activity of lactobacilli immobilized on CbRH increases for different test cultures of Enterobacteriaceae by 25-60 %. As mentioned above, carbonized nanostructured sorbents have a high specific surface area, which, in turn, determines an increase in the adsorption capacity of CbRH [52]. The authors determined the sorption activity of SDSs with respect to LPS (Fig. 18). The concentration of lipopolysaccharide in the medium was determined photometrically on microplates using a QCL-1000® Chromogenic LAL Endpoint Assay (Lonza Group Ltd., Switzerland) and a photometric scanner (Bio-Rad Co., USA).

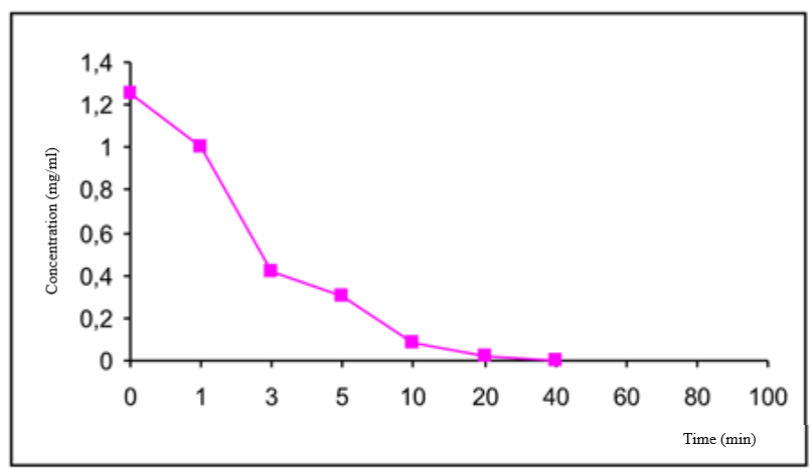

Figure 18. Adsorption of LPS on the CbRH [51].

As can be seen from Figure 18, during the first 10 minutes the LPS binding process is already active, and the concentration of LPS in the solution decreases by 90%. After 40 minutes, this drug is not detected in the solution at all. The authors of this study claim that CbRH is a promising material for toxic shock LPS detoxification [51, 53].

## 3. OBTAINING GRAPHENE STRUCTURES FROM RICE HUSK

Graphene has received much attention because of its unique properties with various expected applications. So far, the methods for growing graphene have mainly been epitaxial growth on silicon carbide [54], graphite crystal layering [55], chemical vapor deposition (CVD)[56] and chemical reduction of graphene oxide [57]. Most graphene synthesis methods require complex multiple steps, either high-temperature or expensive and advanced tools [58]. However, there is stillroom for simpler, more economical and large-scale methods. Recently, a green synthesis process has been introduced to produce graphene from biomass precursors such as chitosan [59], alfalfa plants [60], sugar [61], green tea [62] and other food products [63]. In this review, we will consider the synthesis and characteristics of graphene from agricultural wastes such as Rice husk.

Rice husk, the main by-product of rice milling, are a common form of agricultural waste [64, 65]. In terms of chemical composition, organic components (~ 22% lignin, ~ 38% cellulose, ~ 20% hemicellulose) and inorganic silicon dioxide ($SiO_2$) make up about ~ 20% of the components [66]. Currently, RDF is used as a raw material for biofuel production, power generation, and for fuel boilers [67]. RH and rice husk ash are mainly used to produce silicon and carbon-based materials such as silica [68, 69], silicon (Si) [69], silicon nitride ($Si_3N_4$) [71], silicon tetrachloride [72], silicon carbide [73], zeolites [74], activated carbon [75, 76] and graphene [77, 78]. Graphene derived from rice husk has a unique structure with clean edges, nanoscale holes and topological defects in the carbon lattice, which can cause new physical and chemical properties. Rice husk graphene is expected to offer opportunities for the development of various applications through inexpensive, simple and scalable fabrication.

Therefore, in [79] a multilayer graphene was synthesized using rice husk ash as a promising material for energy storage (Fig. 19). This methodology demonstrates the usefulness of rice husk ash as a carbon source for graphene synthesis and as a protective barrier against oxidation of the original rice husk and KOH mixture. According to the authors, oxidation can occur during the synthesis process due to high-temperature annealing of the rice husk ash and KOH mixture. The electrochemical characteristics showed a decent capacitance value of 86 $F·g^{-1}$ at 500 $mV·s^{-1}$.

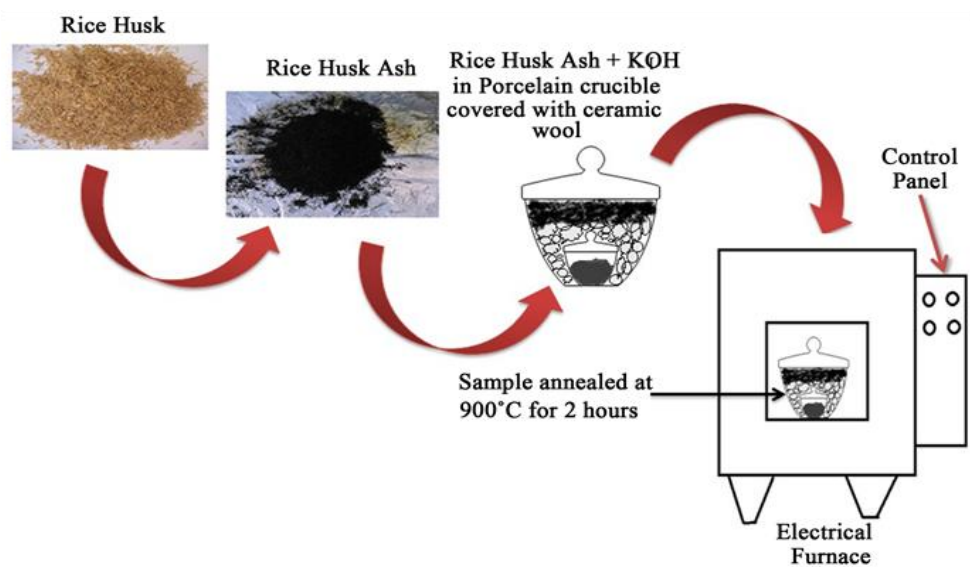

Figure 19. Scheme of graphene synthesis from rice husk ash [79].

The presence of graphite structure in the obtained samples was confirmed by X-ray diffraction analysis, Raman spectroscopy and transmission electron microscopy (Fig. 20). Several layers of graphene and agglomeration of silica particles can be observed in these images. The inset shows an electronogram of a selected region (SAED) where individual spots have merged into rings. This shows the characteristic of a polycrystalline sample and suggests overlapping sheets of graphene and aggregation of silica particles with a random arrangement.

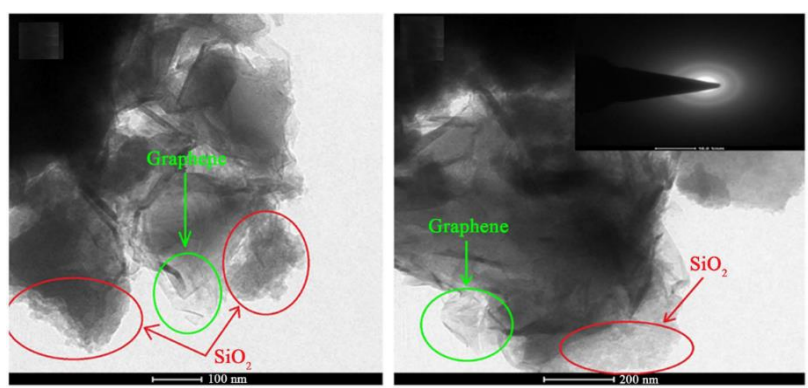

Figure 20. TEM image of graphene obtained from rice husk ash [79].

The novelty of this synthesis method can be characterised as a single chemical synthesis method. The use of a natural precursor makes this method very cost-effective for large-scale production.

In [80], graphene nanosheets were synthesized from brown RH (collected in Tamil Nadu, India), whose electrochemical properties were investigated for their further application as an electrode material. KOH was also used to activate the brown RH. In general, the chemical reactions with the formation of graphene nanosheets occurring during high-temperature annealing of RH can be described as follows:

$$6KOH + 2C\ (Ash) \xrightarrow{700\ °C} 2K + 3H_2 + 2K_2CO_3 \qquad (1)$$

$$K_2CO_3 + SiO_2\ (Ash) \xrightarrow{700\ °C} C + K_2SiO_3 + O_2 \qquad (2)$$

$$K_2SiO_3 + C + H_2O \xrightarrow{RT} C + (K_2SiO_3 + H_2O)\ solution \qquad (3)$$

The new fabrication method developed is quite simple and produces crystalline ultra-thin graphene nanosheets with a low defect density. This is presumably because each fabrication step is optimised. The graphene nanosheets exhibited a crumpled silk-wave plate structure with a large surface area (~1225 $m^2 \cdot g^{-1}$) and high porosity. The electrode from such graphene nanosheet showed excellent specific capacitance (115 $F \cdot g^{-1}$ at 0.5 $mA \cdot cm^{-2}$).

Materials based on graphene can also be used in catalysis. Thus, the paper [81] reports the production of a new nanocomposite from the rice husk ash - reduced graphene oxide by the hydrothermal method and its catalytic application in the Biginelli reaction [82]. It was found that a cost-effective high-performance catalyst based on graphene oxide with reduced rice husk ash can be reused up to 7 repeated cycles with only about 4% decrease of its initial activity.

Another important application of graphene-based materials is the sorption process. Because the two-dimensional structure of graphene due to its selective permeability and mechanical strength allows its use as a sorbent. For example, a group of scientists in their study [83] obtained a nanobiosorbent on the basis of graphene quantum dots from RH with an increased number of surface hydroxyl groups to remove Pb (II) and La (III) from aqueous solutions under the influence of microwave radiation. The study of sorption properties of the obtained material showed that the degree of purification of water samples from lead and lanthanum was 98.5-99.8% and 94.6-96.2% respectively.

Many properties of nanomaterials depend on particle size, so particle analysis is very important. In fields such as nano-medicine, nano-energy, particle size plays an important role [84]. Taguchi statistical tool [85] is used to investigate such parameters. In the work of scientists [86], the preparation and characterisation of graphene from rice husk is carried out using a microwave process and Taguchi tool is used to develop a model for the input and output parameters used in this study.

Graphene from Rice husk was obtained using a microwave process where $Fe(C_5H_5)_2$ ferrocene was used as a catalyst. From the parametric study, it is clear that the influence of temperature for the proposed model is greater than that of the other two variables. The characterisation was done by different methods such as XRD, Zeta-Sizer, FTIR, UV-vis and FESEM-EDX. The average crystallite size of 58.01 nm was obtained from the XRD analysis and the minimum size of 42.5 nm was calculated from the Zeta-sizer method. The authors confirm that the minimum particle size is reached with 30 g ferrocene, 60 g rice husk and a temperature of 750°C. The preparation of graphene has been successfully achieved with this method. Finally, an empirical relationship based on Taguchi's design followed by an approach to historical data for the particle size of graphene produced by a simple microwave process has been developed.

In morphological analysis, flakes of graphene sheets with silica nanoparticles can be seen in figure 20. The prepared graphene has a slightly irregular spherical shape with pores and a high degree of accumulation. Quantitative analysis by EDX showed the presence of carbon, oxygen and silica as the atomic % of carbon, oxygen and silica are respectively 61.73, 30.57 and 7.71%. The results confirm the high purity of graphene.

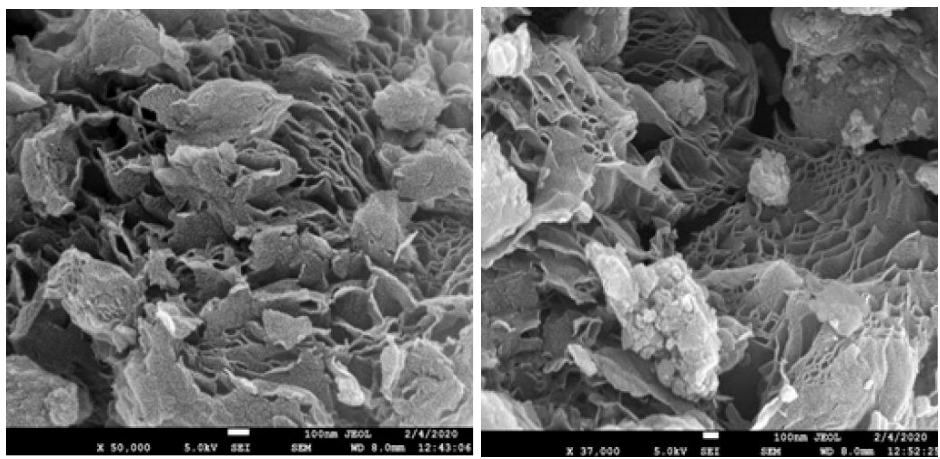

Figure 21. AFEM results for graphene particles [87].

Methylene blue (MB) is a cationic dye commonly used in the textile, paper, plastic, leather and food industries [87]. Wastewater containing MB can adversely affect the health of aquatic life [88]. Conventional methods of dye removal from wastewater include biodegradation, oxidation, ion exchange, precipitation and adsorption [89]. Adsorption is the most widely used method because of its high efficiency, low cost and ease of operation [90]. Thus, a group of scientists synthesized carbon nanocomposites based on graphene oxide [91] to use it as a MS sorbent.

This paper reports the synthesis of carbon materials in the form of graphene oxide composites from rice husk obtained by activating it with $H_3PO_4$ and $ZnCl_2$. These composites were investigated using Raman spectrometer, Fourier transform infrared spectrometer, transmission electron microscope, auto emission scanning electron microscope, X-ray diffractometer and surface area analyser. Experimental results showed that $H_3PO_4$ activated graphene oxide carbon, $ZnCl_2$ activated carbon and ordered mesoporous carbon had a surface area of 361, 732 and 936 $m^2 \cdot g^{-1}$ respectively; pore volume of 0.299, 0.581 and 1.077 $cm^3 \cdot g^{-1}$ respectively; and mean pore size of 2.31, 3.17 and 4.35 nm respectively. Carbon composites with graphene oxide showed higher adsorption capacity than pure carbon materials without graphene oxide. The maximum adsorption capacity using methylene blue as adsorbate corresponded to the order of ordered mesoporous carbon (1591 $mg \cdot g^{-1}$) > activated $ZnCl_2$ carbon (899 $mg \cdot g^{-1}$) > activated $H_3PO_4$ carbon (747 $mg \cdot g^{-1}$). The results of the study show that rice husk wastes have excellent potential to produce very valuable nanoproducts and reduce environmental pollution.

In this work, hybrid nanocomposites based on rice husk graphene (GRHA) and zeolite imidazolate framework-8 (ZIF-8) were prepared for hydrogen adsorption. The main contribution of this work is the simplification of the synthesized GRHA/ZIF-8 hybrid nanocomposites. In addition, the use of synthesized graphene from rice husk (RH) can help in solving environmental problems, as the removal of RH is quite challenging. GRHA was obtained by calcining rice husk ash (RHA) at 900°C for 2 hours in a muffle furnace under atmospheric conditions, while the GRHA/ZIF-8 nanocomposite was obtained under free solvent conditions using deionized water at room temperature for only 1 hour. Adsorption-desorption of $N_2$ indicated a type I isotherm. Interestingly, it was found that the BET (BETSSA) specific surface area of GRHA / ZIF-8 showed an increase of up to 3 times that of the original GRHA with the addition of 0.2 g GRHA. According to the experimental data, the $H_2$ adsorption by GRHA/ZIF-8 nanocomposite obeys the pseudo-second order kinetic and intraparticle diffusion model with $R_2$ values up to 0.9890 and 0.8087 respectively at 12 bar. Moreover, GRHA/ZIF-8 had the highest hydrogen adsorption of 31.84 $mmol \cdot g^{-1}$ at 12 bar. These impressive results are justified by the combination of the microporosity of ZIF-8 and the mesoporosity of GRHA.

Table 3 - Variety of graphene structures derived from RH and their applications

| The resulting material | Temperature / | Applications | Reference |
|---|---|---|---|

|  | combustion time RH |  |  |
|---|---|---|---|
| Multilayer graphene | 900 °C / 2 h | Energy storage | [78] |
| Graphene nanosheets | 500 °C / 2 h | Electrode | [79] |
| Nanocomposite based on reduced graphene oxide | 650 °C / 3 h | Catalysis | [80] |
| Graphene quantum dots | 700 °C / 2 h | Sorption | [82] |
| Graphene oxide-based carbon nanocomposites | 700 °C / 30 min | Sorption | [90] |

**CONCLUSION**

Porous carbon materials from plant raw materials play an important role in a wide range of applications, such as adsorption materials for various systems, as a catalytic carrier, battery and capacitive electrodes, capacitors of various types and gas storage.

Rice husk have traditionally been considered an inexpensive waste product. However, increasingly inflexible regulations governing their disposal have led to scientific efforts to identify potential applications for these materials, such as energy production, pollution control, and building materials, among others, have been identified and developed. However, these materials have other potential applications, and as science and technology advance, many more applications are expected to be found in the future as society increasingly focuses on sustainability and recycling of plant materials.

This review presents current advances in the processing of Rice husk as a promising material for producing activated carbon with a developed specific surface area. The article shows promising applications of porous carbon material from Rice husk as an electrode material for supercapacitors and lithium-sulfur batteries, for water purification from organic and inorganic pollutants, for water desalination and extraction of noble metals from their solutions. It is shown that Rice husk can be a promising material for obtaining graphene-like carbon materials in large-scale production.

A separate section presents the work of the Institute of Combustion Problems in the production of activated carbons and sorbents based on Rice husk. These sorbents were used for selective extraction of noble metals, sorption of metal ions, free hemoglobin and lipopolysaccharide. Antimicrobial properties of sorbents immobilized by different strains of bacteria were studied. The prospects of using rice hulls for obtaining carbon sorbents, activated carbons and their further application in various fields of science and technology have been shown.

This review also included an extensive list of literature on Rice husk, including works by researchers from Kazakhstan. Based on the analyzed literature and suggestions for future research.

**ACKNOWLEDGEMENT**

This research was supported by the Science Committee of the Ministry of Education and Science of the Republic of Kazakhstan (Grant No. OR11465430).